# Tunable magnetism in ferroelectric α-In$_2$Se$_3$ by hole-doping

Chang Liu,[1] Bing Wang,[1] Guanwei Jia,[1] Pengyu Liu,[1] Huabing Yin,[1, a)] Shan Guan,[2, a)] and Zhenxiang Cheng[3]

[1] *Institute for Computational Materials Science, School of Physics and Electronics, International Joint Research Laboratory of New Energy Materials and Devices of Henan Province, Henan University, Kaifeng 475004, China*

[2] *State Key Laboratory of Superlattices and Microstructures, Institute of Semiconductors, Chinese Academy of Sciences, Beijing 100083, China*

[3] *Institute for Superconducting and Electronic Materials, University of Wollongong, North Wollonggong, NSW 2500, Australia*

Two-dimensional (2D) multiferroics attract intensive investigations because of underlying science and their potential applications. Although many 2D systems have been observed/predicted to be ferroelectric or ferromagnetic, 2D materials with both ferroic properties are still scarce. By using first-principles calculations, we predict that hole-doping can induce robust ferromagnetism in 2D ferroelectric α-In$_2$Se$_3$ due to its unique flat band structure, and the Curie temperature ($T_C$) can be much higher than room temperature. Moreover, the doping concentration, strain, and number of layers can effectively modulate the magnetic moment and the $T_C$ of the material. Interestingly, strong magnetoelectric coupling is found at the surface of hole doped multilayer α-In$_2$Se$_3$, which allows non-volatile electric control of magnetization. Our work provides a feasible approach for designing/searching 2D multiferroics with great potential in future device applications, such as memory devices and sensors.

---

[a)] Author to whom correspondence should be addressed: yhb@henu.edu.cn; shan_guan@semi.ac.cn

Multiferroics are a special class of materials which simultaneously exhibit more than one ferroic ordering, such as ferromagnetism (FM), ferroelectricity (FE), and ferroelasticity. Multiferroics with coexisting FM and FE hold great promise in non-volatile storage devices due to the inherent coupling between the two ferroic orderings.[1-4] Previous researches are mainly focused on the three-dimensional (3D) multiferroics with perovskite structures, such as $BiFeO_3$ and $TbMnO_3$.[1, 2] However, when the multiferroics film thickness is reduced to a few nanometers, surface effect such as dangling bonds will seriously affect the performance of materials. Moreover, due to the finite size-effect and depolarization electrostatic field, conventional ferroelectric materials usually cannot maintain the polarization below a critical thickness,[5, 6] and the long-range magnetic ordering, as demonstrated by the Mermin–Wagner theorem, is strongly suppressed by thermal fluctuations in ultrathin isotropic films.[7] This thickness effect greatly limits the application of 3D multiferroics in future microelectronic devices which need ultrathin films as the building components.

With the discovery of graphene,[8] 2D materials have become an emerging hot research area due to their interesting properties and promising applications. With the continuous efforts of scientific researchers, more and more 2D materials have been discovered, such as monoelement 2D materials (i.e., silicene, phosphorene, etc.),[9] transition metal dichalcogenides,[10] hexagonal boron nitride,[11] and Mxenes.[12] Their various physical properties such as conductivity, flexibility, optical properties, catalytic properties and electronic structures have also been extensively studied.[13-16] Recently, both ferroelectricity and ferromagnetism have been discovered in 2D materials. For example, 1 T-$MoS_2$,[17] group-IV or III–V honeycomb binary compounds,[18] IV-VI and group-V binary compounds,[19-24] triple-layer $LaOBiS_2$,[25] β-GeSe,[26] and β-GeS[27] have been theoretically predicted to be 2D FE materials. While several 2D FE materials such as SnTe,[28] $CuInP_2S_6$,[29] and $In_2Se_3$[30, 31] have been synthesized experimentally. On the other hand, more than 40 intrinsic 2D FM materials have been predicted theoretically within the framework of density functional theory,[32, 33] such as Mn- and Cr-based semiconducting trichalcogenides,[34] $CrXTe_3$ (X = Si, Ge) nanosheets,[35] etc.[36, 37] Some of them have been experimentally realized, such as bilayer $Gr_2Ge_2Te_6$,[38] monolayer chromium trihalides ($CrX_3$),[39-41] and $VSe_2$.[42] Although many 2D FE and 2D FM materials have been reported, 2D multiferroics with simultaneous ferroelectricity and ferromagnetism are still rare. Only very limited number of 2D systems have been theoretically predicted to be multiferroics, such as binary graphitic bilayer,[43] 2D CrN,[44] transition-metal halide monolayer,[45] etc.[46-49] Our previous work proposed a material γ-GeSe that can achieve 2D multiferrocity through carrier doping.[49] However, it is diffcult to achieve strong magnetoelectric coupling in this material, and it has not been synthesized experimentally.

In this work, based on the first-principles calculation, we reveal the doping-induced ferromagnetism in experimentally synthesized 2D FE compound α-$In_2Se_3$.[30] We find that hole-doping in a large concentration range ($1.4 \times 10^{14}/cm^2 \sim 29 \times 10^{14}/cm^2$) can introduce robust ferromagnetism in α-$In_2Se_3$ with a $T_C$ much higher than room temperature. By applying strain or changing the doping level, the magnetic moment, $T_C$, and the spin polarization energy can be widely modulated. Particularly, this hole doping-induced ferromagnetism can also exist in multilayer form of α-$In_2Se_3$. Furthermore, hole-doped multilayer α-$In_2Se_3$ shows strong magnetoelectric coupling effect, where the magnetic moment of the surface layer can be tuned by switching of the spontaneous polarization. The multi-mode controllable ferromagnetism, and high T$_C$ in α-In$_2$Se$_3$ provide a promising material platform for future applications such as memory and sensor devices with high density, low energy consumption and fast speed.

All the calculations were performed using the Vienna Ab initio Simulation Package (VASP).[50] The projector augmented wave (PAW) method[51] for treating ion-electron interactions is used, and the Perdew-Burke-Ernzerhof (PBE) type[52] generalized gradient approximation (GGA) is used to describe the exchange and correlation interaction. After the convergence tests, we use 450 eV as the cutoff energy for the plane wave basis and 17×17×1 as the Monkhorst-Pack k mesh for the Brillouin-zone (BZ) sampling. The atomic forces convergence criterion was set to 0.01 eV/Å for the optimized structures. A vacuum region of 15 Å was added in the lattice to avoid the interactions between different layers. The Heyd–Scuseria–Ernzerhof (HSE) hybrid functional[53] was also used in the band structure calculations. The optB88-vdW functional[54] are used to describe the interlayer interactions in multilayer In$_2$Se$_3$. The effect of charge doping is simulated by changing the total number of electrons in the unit cell, with a compensating jellium background of opposite charges to maintain neutrality. In the estimation of Curie temperature, the finite temperature was introduced by changing the smearing factor ($\sigma = k_B T$) in the Fermi–Dirac distribution function: $f\left(\frac{\epsilon-\mu}{\sigma}\right) = \frac{1}{\exp\left(\frac{\epsilon-\mu}{\sigma}\right)+1}$.[55]

An monolayer α-In$_2$Se$_3$ layer contains five atomic layers in the sequence of Se–In–Se–In–Se. The lattice constant of the optimized structure is 4.107 Å, which is in consistent with previous theoretical work.[56] From Fig. 1 (a) we can see that there are four different bond lengths between In and Se ions, which are 2.72 Å, 2.91 Å, 2.56 Å, and 2.68 Å, respectively. The upper In ion in Fig.1 (a) has 8 coordinated Se ions, forming an octahedral structure, while the lower In ion forms a tetrahedral coordination. The asymmetric positions of Se ions in the middle atomic layer lead to spontaneous polarization, which enables α-In$_2$Se$_3$ a 2D ferroelectric material. The electronic structures calculated from both the PBE functional and the HSE06 method show that the two band structures have similar dispersions except for the different indirect band gap values, which are approximately 0.77 eV and 1.61 eV (Fig. 1 (c) and (d)), respectively. The valence band maximum (VBM) is located on the M–Γ line, and the conduction band minimum (CBM) is located at the Γ point. From the projected density of states (PDOS) one can find that the DOS around VBM is mainly contributed by Se-4p states, and the CBM is mainly composted of both In-5p and Se-4p states. It is worth noting that the DOS increases suddenly at the VBM, this is caused by the flat valence band of the highest occupied state, which is clearly shown in the 3D projection band in Fig. 1 (b). The flat band near the Fermi level usually causes electronic instability such as itinerant ferromagnetism induced by doping carriers in nonmagnetic system.[57, 58]

Guided by this idea, we introduce holes into this valence band and explore possible ferromagnetism occurs in monolayer α-In$_2$Se$_3$. As shown in Fig. 2 (a), one observes that, as the hole doping level gradually increases, the magnetic moment suddenly emerges at a concentration of about 0.8×10$^{14}$/cm$^2$ and rises rapidly to about 1 μ$_B$/hole. The ferromagnetism can be maintained in a large range of doping concentrations, and finally vanish at a rather high doping concentration of about 29×10$^{14}$/cm$^2$. The spin polarization energy (defined as the energy difference between the nonmagnetic state and the FM state) is calculated to further confirm the doping concentration range for ferromagnetism emerging (see Fig. 2 (b)). One finds that the spin polarization energy remains positive values in a large hole concentration range ($1.4\times10^{14}$/cm$^2$ ~ $29\times10^{14}$/cm$^2$), implying that the ground state of α-In$_2$Se$_3$ is FM in this doping range.

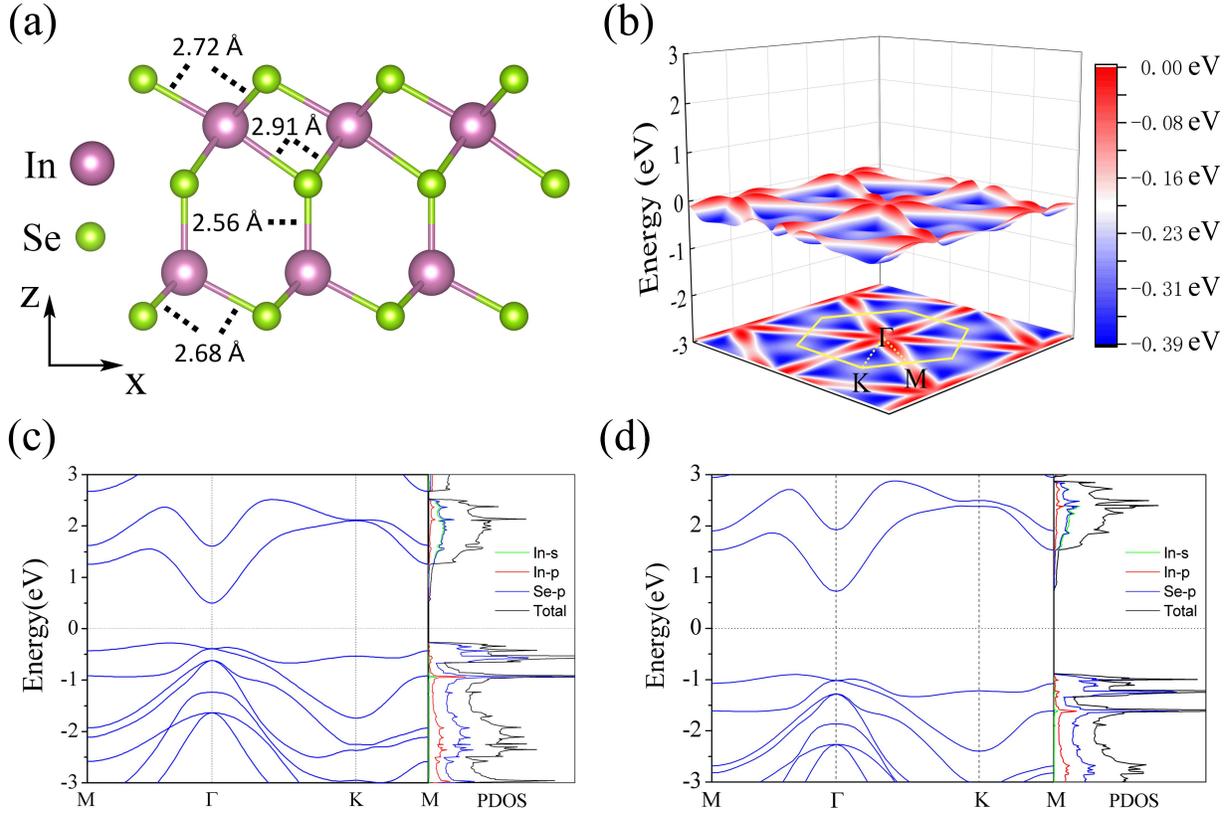

Fig. 1. (a) Lattice structure of monolayer α-In$_2$Se$_3$; (b) 3D projection of the flat valence band; (c) and (d) Band structures and PDOS of monolayer α-In$_2$Se$_3$ calculated by using the PBE functional and the HSE06 method, respectively.

Fig. 2 (c) illustrates the spin-polarized band structure when the doping concentration is n = $1.78\times10^{14}$/cm$^2$. We can see that the spin degeneracy is broken with a large exchange splitting of approximate 0.15 eV. The spin-up band (blue line) is fully occupied, whereas the spin-down band (red line) crosses the Fermi Level. Consequently, all the holes are doped into the spin-down band, leaving this band partially filled. As a result, hole-doping makes α-In$_2$Se$_3$ a 2D half metal, demonstrating an ideal candidate for fabricating spintronic devices.

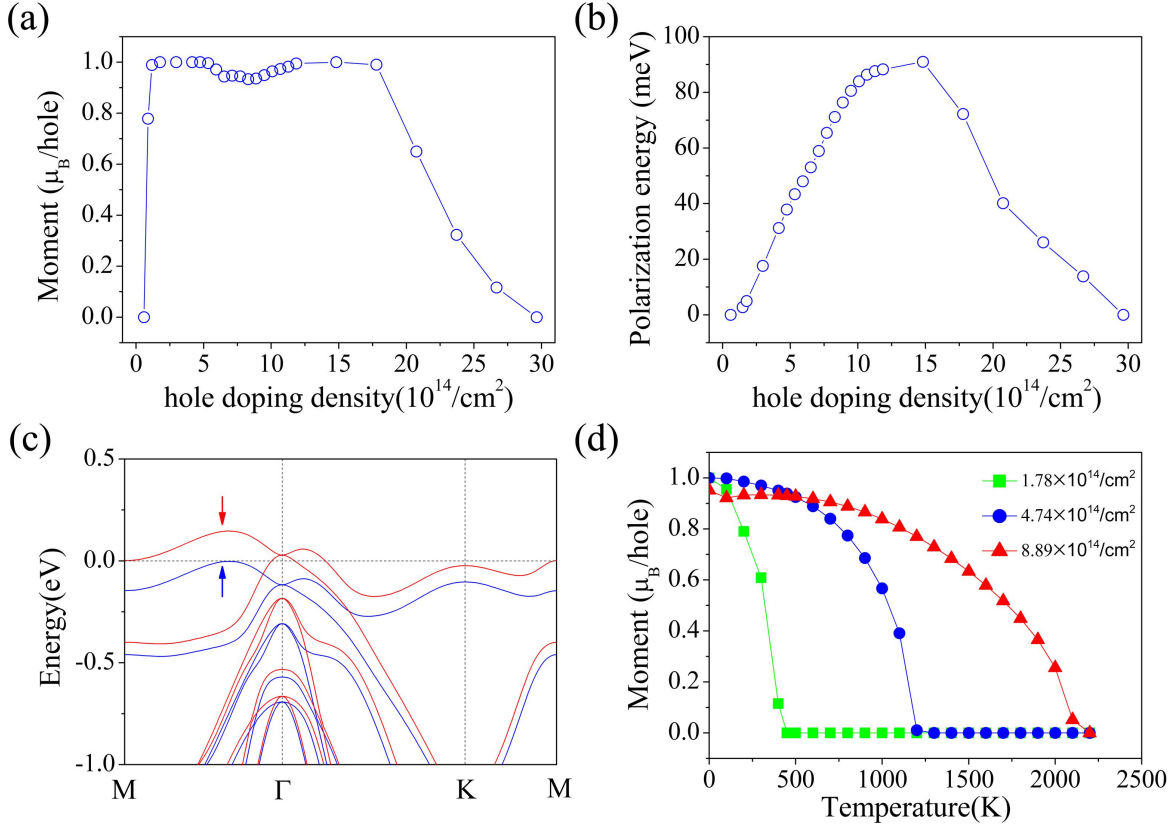

Fig. 2. (a) Doping concentration dependence of the magnetic moment per doping hole; (b) Doping concentration dependence of the spin polarization energy; (c) Spin-polarized band structure in the doping concentration of $1.78\times10^{14}/cm^2$; (d) Doping-induced magnetic moment vs temperature for three different doping levels.

High $T_C$ is highly desired for the application of magnetic materials. However, materials with robust ferromagnetism at room temperature are scarce. In order to study the thermodynamic stability of the itinerant ferromagnetism induced by doping, we choose three different doping levels and show how the magnetic moment of doped holes varies with temperature. As depicted in Fig. 2 (d), when the doping level is relatively low (n = $1.78\times10^{14}/cm^2$), the doping-induced magnetic moment is suppressed at about 450 K. Fortunately, higher doping concentration can significantly increase the ferromagnetic phase transition temperature $T_C$. For instance, the $T_C$ rise up to about 1200 K and 2100 K at n = $4.74\times10^{14}/cm^2$, and n = $8.89\times10^{14}/cm^2$, respectively. Remarkably, such a high Curie temperature is not only higher than the predicted intrinsic ferromagnetic materials,[36] but also higher than other 2D materials with doping-induced ferromagnetism.[49, 57, 58]

The origin of doping-induced magnetism can be qualitatively described by the Stoner model.[57] That is to say the itinerant ferromagnetism will probably occur once the criterion $ID(E_f) > 1$ is satisfied, where $I$ is the strength of the exchange interaction that can be estimated from the spin-splitting in the FM state, and $D(E_f)$ is the (non-spin polarized) DOS at the Fermi level. For n = $1.78 \times 10^{14}$/cm$^2$, the Stoner factor I and $D(E_f)$ are calculated to be 0.5 and 5.0, respectively. These values clearly satisfy the Stoner criterion $ID(E_f) > 1$; thus, the itinerant ferromagnetism emerges in hole-doped monolayer α-In$_2$Se$_3$.

2D materials usually possess excellent flexibility and can sustain much larger strains than their bulk crystals,[59] offering strain as an easy-to-implement mean to modulate material properties. For monolayer α-In$_2$Se$_3$ with different doping levels, we apply biaxial strains to investigate the effect on the doping-induced ferromagnetism. From Fig. 3 (a) one can see that tensile strain can enhance the ferromagnetism. For the doping concentration n ≥ $1.78 \times 10^{14}$/cm$^2$, the magnetism is already saturated with a moment of 1 μ$_B$/hole without strain, so tensile strain cannot enhance the magnetization. In comparison, compressive strains tend to suppress the doping-induced magnetic moment. For example, for n = $1.78 \times 10^{14}$/cm$^2$, a compressive strain of 3% can reduce the magnetic moment to nearly zero. Moreover, low doping level, such as n = $0.3 \times 10^{14}$/cm$^2$, is not sufficient to induce ferromagnetism in unstrained α-In$_2$Se$_3$, but a small tensile strain of 3% will produce a large magnetic moment of about 0.9 μ$_B$/hole, driving α-In$_2$Se$_3$ from paramagnetic state to ferromagnetic state. The strain-induced magnetic phase transition indicates the existence of strong coupling between mechanical strain and magnetism, suggesting the huge potential of hole-doped In$_2$Se$_3$ for realization of multi-fields coupling for sensors.

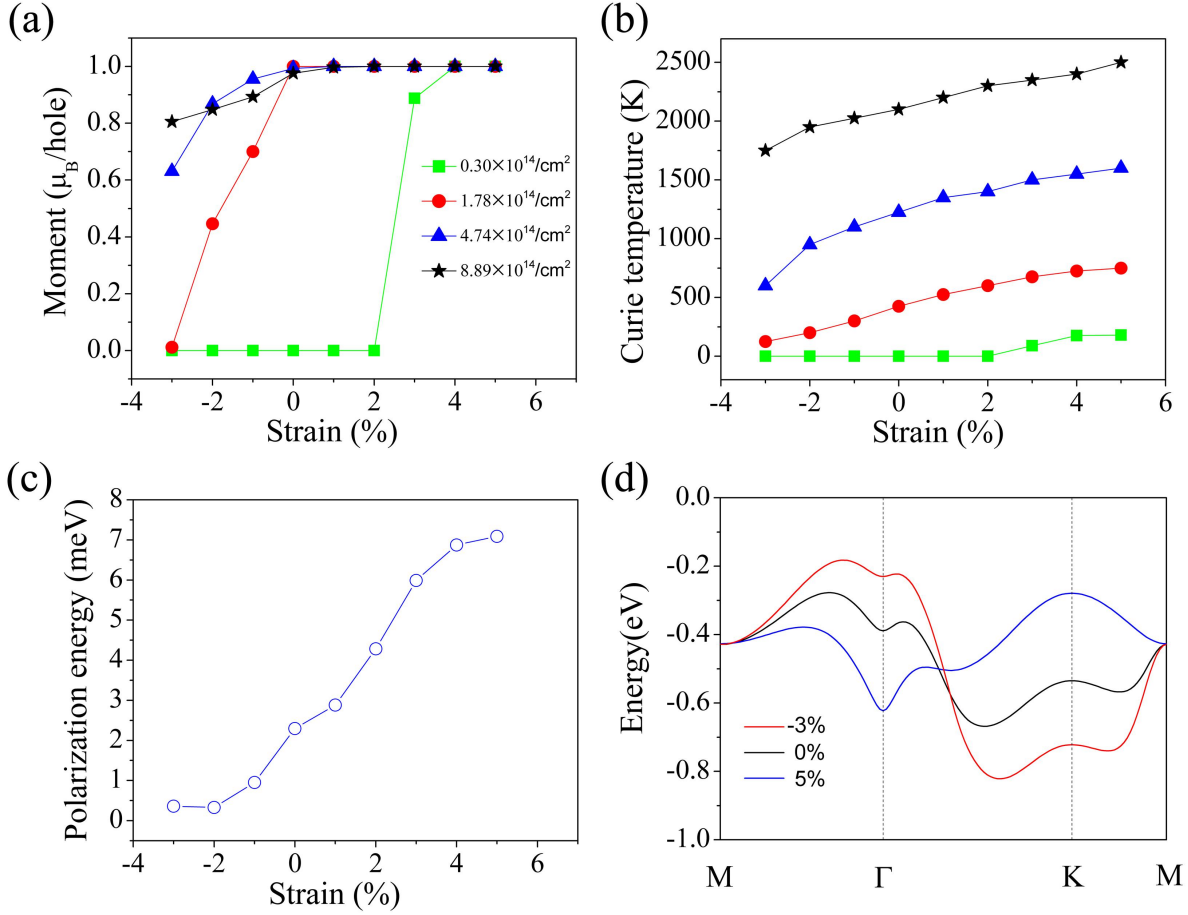

Fig. 3. Effect of strain on (a) magnetic moment for different doping concentrations, (b) on $T_C$, and (c) on spin polarization energy. (d) The uppermost valence band of monolayer α-In$_2$Se$_3$ under 3% compression strain (red line), no strain (black line), and 5% tensile strain (blue line), respectively.

The $T_C$ can also be well modulated by strain. Similar to the effect of strain on magnetic moment, $T_C$ can be enhanced by a tensile strain but be suppressed by a compressive strain (see Fig. 3 (b)). In detail, for n = $0.3\times10^{14}$/cm$^2$, $1.78\times10^{14}$/cm$^2$, $4.74\times10^{14}$/cm$^2$ and $8.89\times10^{14}$/cm$^2$, -3%~5% biaxial strain can modulate $T_C$ in the range of 0 K~180 K, 125 K~750 K, 600 K~1600 K and 1750 K~2500 K, respectively. In Fig. 3 (c), we plot the dependence of spin polarization energy on strain for n = $1.78\times10^{14}$/cm$^2$ as an example. The energy-strain curve also shows a similar tendency to the moment-strain and $T_C$-strain curves. In general, higher spin polarization energy indicates stronger ferromagnetic coupling which leads to higher $T_C$. Therefore, the relationship between strain and spin polarization energy confirms the calculation result of $T_C$-strain curves.

Through electronic structure, we explore why strain can modulate the magnetism in α-In$_2$Se$_3$. Fig. 3 (d) shows how the valence band closest to the Fermi level changes when applying strain. When applying a 3% compressive strain, the bandwidth expanded from 0.39 eV to 0.64 eV, while a 5% tensile strain narrows the bandwidth from 0.39 eV to 0.34 eV. Flatter band lead to enhanced Stoner instability, and therefore stronger ferromagnetism emerges. Conversely, compressive strain shortens the atomic bond length, thereby increasing the bandwidth and weakening the ferromagnetism.

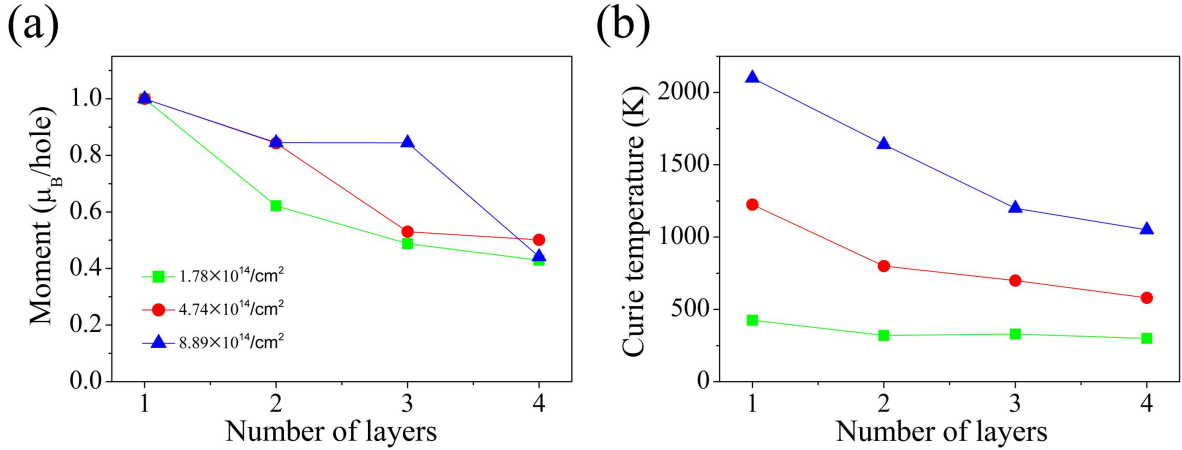

Fig. 4. (a) Doping-induced magnetic moment and (b) T$_C$ as a function of the number of layers under three different doping concentrations.

Does multilayer α-In$_2$Se$_3$ possess doping-induced ferromagnetism? We build 1-layer to 4-layer of In$_2$Se$_3$ multilayers, respectively, and calculate the magnetic moment and the T$_C$ as hole doped. The results are shown in Fig. 4, one observes that both the moment and the T$_C$ decrease with the increasing number of layers. However, even in the 4-layer In$_2$Se$_3$, the doping-induced moment is also sizeable (~ 0.5 μ$_B$/hole), and the T$_C$ is still above room temperature (300 K ~ 1050 K) for n = 1.78×10$^{14}$/cm$^2$ ~ 8.89×10$^{14}$/cm$^2$. So we can say that the ferromagnetism is robust at least below 4-layers. Our calculation thus confirms that doping-induced ferromagnetism does retain in the multilayer In$_2$Se$_3$; additionally, it provides another method to tailor the magnetic properties by tuning the thickness of In$_2$Se$_3$ film.

We then take 3-layer In$_2$Se$_3$ with hole concentration n = 4.74×10$^{14}$/cm$^2$ as an example to study strain effects in multilayers. The results are shown in Fig. 5 (a) and (b). One finds that a small tensile strain (~ 1%) can effectively increase the magnetic moment from ~ 0.5 to 1 μ$_B$/hole, while a small compressive strain (~ 2%) can quench the moment, which is similar to that in monolayer. Correspondingly, applying strain can also effectively modulate the T$_C$, with T$_C$ being reduced to zero only via a small compressive strain.

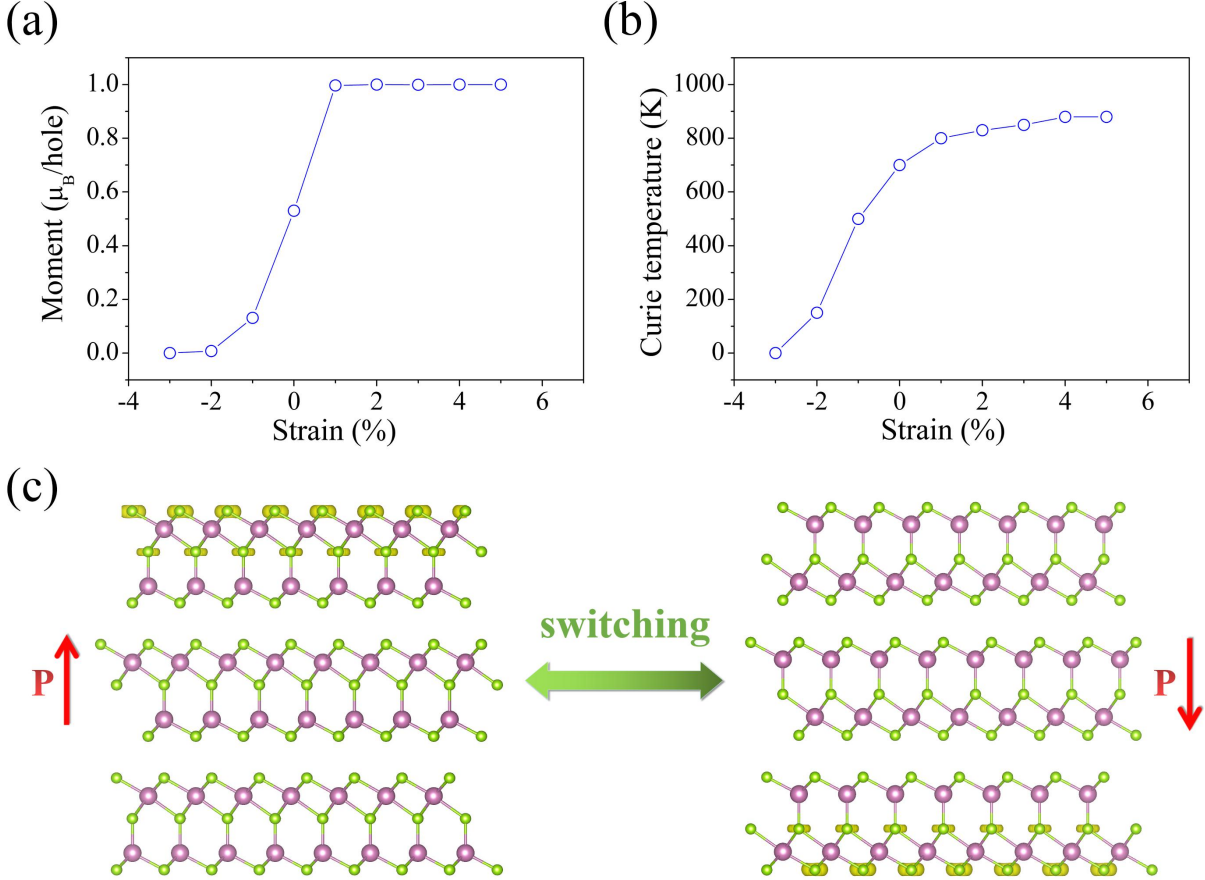

Fig. 5. Strain dependence of (a) magnetic moment and (b) $T_C$ in three-layer $In_2Se_3$ with hole concentration n = $4.74\times10^{14}/cm^2$; (c) Spin-polarized charge density of three-layer $In^2Se^3$ under different electric polarization directions.

Realization of the electrically control of magnetization is highly desirable for novel applications of multiferroic materials. Through the calculation of spin-polarized charge density, we show that hole-doped multilayer $In_2Se_3$ possesses magnetoelectric coupling, which provides an effective route towards the tunable surface magnetic polarization via external electric fields. As shown in Fig. 5 (c), the majority spin is mainly distributed on the upper surface Se ions of the multilayer film. This is because the out-of-plane ferroelectric polarization in $In_2Se_3$ generates a built-in electric field, driving magnetic hole carriers to migrate to the surface along the polarization direction. Therefore, one may naturally expect that the surface magnetic moment can be tuned by applying an electric field to switch the direction of ferroelectric polarization, which is similar to the carrier-mediated magnetoelectric coupling mechanism in the magnetoelectric heterostructures.[60] Taking three-layer $In_2Se_3$ with n = $4.74\times10^{14}/cm^2$ as an example, for one surface, the magnetic moment is about 0.24 $\mu_B$/unit cell, and for the other surface, the magnetic moment is about 0.09 $\mu_B$/unit cell. Therefore, a change in magnetic moment of 0.15 $\mu_B$/unit cell can be achieved by polarization switching, which is comparable to the electric field induced moment variation in composite multiferroic heterostructures with strong magnetoelectric coupling.[61] Moreover, the magnetic moment at the surface is easier to detect than at the interface, making In$_2$Se$_3$ better than the magnetoelectric heterostructures for spintronics devices.

The doping density on the order of $10^{14}$/cm$^2$ can be achieved via the available gating technique, which has been demonstrated in graphene.[62, 63] Therefore, the hole-doped ferromagnetism may be realized experimentally, and make α-In$_2$Se$_3$ a 2D multiferroic material with strong magnetoelectric coupling. One might argue that hole doping makes α-In$_2$Se a half-metal with a strong screening effect; therefore, it is difficult to switch the ferroelectric polarization by applying the electric field. However, doping only introduces few carriers, which is not enough to screen the external electric field. In addition, the electrons are confined within the slab and hardly to conduct along the out-of-plane direction in 2D polar metal.[44] Moreover, besides applying electric field, there are some other method can also effectively switch the ferroelectric domain, such as applying mechanical force[64, 65] or using laser pulses.[66] Hole-doped α-In$_2$Se$_3$ thus has both switchable electric polarization and magnetic polarization, serving as a promising material platform to explore the intriguing physics of 2D multiferroics.

In summary, by using first-principles calculations, we propose that hole doping can induce robust ferromagnetism in α-In$_2$Se$_3$, and turn α-In$_2$Se$_3$ to a 2D multiferroic material. Both the magnetic moment and the T$_C$ are dependent on the doping concentration, and can be well modulated by means of strain. The ferromagnetism of the multilayer In$_2$Se$_3$ is slightly weaker than that of the monolayer, but a small tensile strain can significantly increase the magnetization and T$_C$. More importantly, due to the built-in electric field, the magnetic moment in the multilayer α-In$_2$Se$_3$ is mainly distributed on the surface and can be effectively controlled by electric polarization reversal. These findings reveal the application potential of α-In$_2$Se$_3$ for fabricating tunable nano-spintronics devices.

This work was financially supported by the National Key Research and Development Program of China (Grant No. 2018YFB2202800), the National Natural Science Foundation of China (Grant No. 11904079), the Natural Science Foundation of Henan (Grant No. 202300410069), and the China Postdoctoral Science Foundation (Grants No. 2019M652303, 2020M682274, and 2020TQ0089).

**Data Availability Statements**

The data that supports the findings of this study are available within the article.